\documentclass[journal,preprint]{vgtc}       

\ifpdf                  
  \pdfoutput=1\relax                 
  \usepackage{graphicx}
  \DeclareGraphicsExtensions{.pdf,.png,.jpg,.jpeg}
\else
  \usepackage{graphicx}
  \DeclareGraphicsExtensions{.eps}
\fi

\graphicspath{{figures/}{pictures/}{images/}{./}} 

\usepackage{microtype}
\PassOptionsToPackage{warn}{textcomp}
\usepackage{textcomp}
\usepackage{mathptmx}
\usepackage{times}

\usepackage{cite}                      
\usepackage{tabu}               
\usepackage{booktabs}

\usepackage[mathscr]{euscript}
\usepackage{amsbsy}
\usepackage{amsfonts}

\usepackage{paralist}
\usepackage[font={small,sf},caption=false,captionskip=1pt]{subfig}
\usepackage{color, soul}
\usepackage[font={scriptsize,sf},tableposition=top]{caption}
\usepackage{float}
\floatstyle{plaintop}
\restylefloat{table}
\usepackage{siunitx}
\usepackage{colortbl}
\usepackage{sparklines}
\usepackage{icomma}

\usepackage{multicol}
\usepackage[bookmarks=false]{hyperref}

\preprinttext{}
\ieeedoi{10.1109/TVCG.2020.3028889}

\onlineid{1254}
\vgtccategory{Research}
\vgtcpapertype{algorithm/technique}

\newcommand{\Title}{
A Visual Analytics Framework for Reviewing \\Multivariate Time-Series Data with Dimensionality Reduction
}
\title{\Title}

\author{Takanori Fujiwara}
\authorfooter{
\item
 Takanori Fujiwara is with University of California, Davis. E-mail: tfujiwara@ucdavis.edu.
}

\author{Takanori Fujiwara, Shilpika, Naohisa Sakamoto, Jorji Nonaka, Keiji Yamamoto, and Kwan-Liu Ma}
\authorfooter{
\item
 Takanori Fujiwara, Shilpika, and Kwan-Liu Ma are with University of California, Davis. E-mail: \{tfujiwara, fshilpika, klma\}@ucdavis.edu.
\item
 Naohisa Sakamoto is with Kobe University. E-mail: naohisa.sakamoto@people.kobe-u.ac.jp.
\item
 Jorji Nonaka and Keiji Yamamoto are with RIKEN R-CCS. E-mail: \{jorji, keiji.yamamoto\}@riken.jp.
}

\shortauthortitle{Fujiwara \MakeLowercase{\textit{et al.}}: \Title}

\newcommand{\name}{MulTiDR}

\newcommand{\Scalar}[1]{$\lowercase{#1}$}
\newcommand{\Vec}[1]{$\mathbf{\lowercase{#1}}$}
\newcommand{\Mat}[1]{$\mathbf{\uppercase{#1}}$}
\newcommand{\Tensor}[1]{$\pmb{\mathscr{\uppercase{#1}}}$}

\newcommand{\TDRView}{TDR view}
\newcommand{\SIView}{SI view}
\newcommand{\FCView}{FC view}
\newcommand{\PMView}{PM view}
\newcommand{\HCView}{HC view}
\newcommand{\FirstDR}{\Mat{Y}}
\newcommand{\SecondDR}{\Mat{Z}}
\newcommand{\ParaMap}{\Vec{w}}
\newcommand{\FC}{\Vec{a}}

\abstract{
Data-driven problem solving in many real-world applications involves analysis of time-dependent multivariate data, for which dimensionality reduction (DR) methods are often used to uncover the intrinsic structure and features of the data. However, DR is usually applied to a subset of data that is either single-time-point multivariate or univariate time-series, resulting in the need to manually examine and correlate the DR results out of different data subsets. When the number of dimensions is large either in terms of the number of time points or attributes, this manual task becomes too tedious and infeasible. In this paper, we present MulTiDR, a new DR framework that enables processing of time-dependent multivariate data as a whole to provide a comprehensive overview of the data. With the framework, we employ DR in two steps. When treating the instances, time points, and attributes of the data as a 3D array, the first DR step reduces the three axes of the array to two, and the second DR step visualizes the data in a lower-dimensional space. In addition, by coupling with a contrastive learning method and interactive visualizations, our framework enhances analysts' ability to interpret DR results. We demonstrate the effectiveness of our framework with four case studies using real-world datasets.
}


\CCScatlist{ 
    \CCScat{I.3.8}{Computer Graphics}{Applications}%
}

\vgtcinsertpkg

\begin{document}
\firstsection{Introduction}
\maketitle

Analysis of multivariate time-series data is becoming increasingly important to studying various phenomena in the real world. 
For example, analyzing electronic health records (EHRs) that contain temporal changes of individuals' various medical measures (e.g., blood pressure and heart rate) for cohort studies can help clinical researchers develop healthcare plans~\cite{lee2017big,huang2019patient,guo2020comparative}.
Many other analysis examples can be found in other domains, such as diagnosis of the performance of parallel computing systems~\cite{muelder2016visual,fujiwara2018visual,guo2018valse}, fault detection of factory assembly lines~\cite{xu2017vidx,wu2018visual,zhou2019survey}, and optimization of transportation systems~\cite{liu2018tpflow,dai2019visual}. 
As seen in the emergence of the Internet of Things, the growing capability and use of sensing devices improves the granularity, quality, and accessibility of multivariate time-series data~\cite{sun2016internet,dimitrov2016medical,ahmed2017role}; at the same time, the increase of the data size and dimensionality makes analysis tasks more challenging~\cite{ahmed2017role}.

To effectively analyze and visualize large, high-dimensional data, dimensionality reduction (DR) methods are often used~\cite{sacha2016visual,liu2016visualizing} because of their ability to provide a succinct overview of such complex data. 
Currently available DR methods designed for 2D or 3D visualization~\cite{van2009dimensionality,cunningham2015linear} can be only applied to data that can be formed into a 2D matrix, such as single-time-point multivariate data (matrix rows: instances, columns: variables), univariate time-series data (rows: instances, columns: time points), and multivariate time-series with a single instance (rows: time points, columns: variables).
When multivariate time-series data consists of multiple instances, the data is often represented as a third-order tensor (or 3D array); consequently, we either cannot directly apply some DR methods such as principal component analysis to the data or we do not know how to properly prepare a distance matrix as an input for other DR methods (e.g., t-SNE~\cite{maaten2008visualizing}).

A common approach to the above problem is slicing the 3D array and then applying a DR method to each resulting slice~\cite{bach2017descriptive}.
For instance, when slicing along a temporal direction, where each slice represents a matrix of instances and variables, we can visualize a set of DR results with animation or small multiples~\cite{bach2017descriptive}. 
However, when a sliced direction has high dimensionality (e.g., 100 time points), the analyst must examine a large amount of DR results and can easily overlook important patterns (e.g., the emerge of outliers). 

To support effective analysis of multivariate time-series data, we introduce a visual analytics framework, \name{}, which employs a two-step DR to generate an overview of the data and supports interpreting the DR results with contrastive learning (CL) and interactive visualization. 
Particularly, in the first step of DR, \name{} compresses and converts a third-order tensor into a matrix, and then, in the second step, it projects high-dimensional data points into a lower-dimensional space.
Similar to the existing DR methods, the two-step DR result shows similarities of instances, variables, or time points and enables visual identification of essential patterns, such as clusters and outliers. 
When compared with ordinary DR, the two-step DR result is derived from two different directions (e.g., variables and time points) and could be more difficult to understand why specific patterns appear.
Thus, to support the analysis of the two-step DR, we integrate CL to identify essential aspects for the analysts to review and interpret in detail with interactive visualization.
We demonstrate the effectiveness of \name{} for multivariate time-series analysis with multiple case studies using real-world datasets and also make qualitative comparisons of \name{} with other potential DR methods.

\section{Background and Related Work}
We provide a brief description of third-order tensors and discuss relevant works.

\subsection{Third-Order Tensors}

A third-order tensor is a 3D array (note that first- and second-order tensors correspond to vectors and matrices, respectively).
Each axis of a tensor is called \textit{mode}.
When a third-order tensor represents multivariate time-series data, the three modes correspond to time points, instances (or samples), and variables (or attributes).
As the main analysis target, we focus on third-order tensors of multivariate time-series data; however, our framework, \name{}, is designed to be able to deal with the other types of third-order tensors.  

The notations used in this paper follow the conventions in the literature~\cite{kolda2009tensor}.
We denote scalars, vectors, matrices, and third-order tensors with lowercase (e.g., \Scalar{x}), boldface lowercase (e.g., \Vec{x}), boldface uppercase (e.g., \Mat{x}), and boldface Euler script (e.g., \Tensor{x}) letters, respectively.
We use indices $t=1, \ldots, T$, $n=1, \ldots, N$, and $d=1, \ldots, D$ for time points, instances, and variables, respectively. 
Here $T$, $N$, and $D$ are lengths of modes of time points, instances, and variables, respectively (i.e., a third-order tensor \Tensor{x} $\in \mathbb{R}^{T \times N \times D}$).

\subsection{Related Work}
\label{sec:related_work}

Our work relates to visual analytics of third-order tensors. 
A third-order tensor commonly found in the visualization field is a ``generalized'' space-time cube~\cite{bach2014review,bach2017descriptive}.
A generalized space-time cube represents a 2D visualization space that changes over time (e.g., temporal geospatial visualizations and animated 2D scatterplots).
Bach et al.~\cite{bach2014review,bach2017descriptive} provided a comprehensive survey of visualizations of generalized space-time cubes.
They also provided a categorization of visualization strategies.
The strategies include 3D rendering (i.e., render a cube as it is), time cutting (i.e., extracting a 2D snapshot at a particular time point), time flattening (i.e., collapsing temporal changes into a single 2D image), time juxtaposing (i.e., arranging multiple 2D snapshots as small multiples), space cutting (i.e., extracting a planar cut in one direction of the 2D space), and among others.
One strategy that the survey did not discuss in detail is dimensionality reduction (DR), which can be considered as a special form of flattening.
Since our framework, \name{}, also employs DR, here we focus on discussing the works using DR to visualize third-order tensors. 

Similar to our work, a target application of many of the existing works is visualizing multivariate time-series data.
One simple strategy is applying DR to a matrix of instances and variables at each time point and then showing temporal changes with animation or juxtaposition. 
To support such a visualization, the researchers developed dynamic DR methods that provide coherent node positions between consecutive time points, such as the time-based least square projection~\cite{alencar2012time}, Dynamic t-SNE~\cite{rauber2016visualizing}, and the enhanced incremental principal component analysis (PCA)~\cite{fujiwara2019incremental}.
However, finding useful patterns, such as outliers or similar time points, is difficult when relying on animation or juxtaposition.

Another common strategy is applying DR based on a dissimilarity of each time point's matrix~\cite{bach2016time,van2016reducing,jackle2016temporal,muelder2016visual,von2015mobilitygraphs,fujiwara2017visual}. 
This strategy generates an overview of the (dis)similarities of time points.
For example, Bach et al.~\cite{bach2016time} computed the dissimilarity of each pair of 2D images at different time points with a certain distance measure, such as a Euclidean distance; then applied multidimensional scaling (MDS) based on their dissimilarities.
In the MDS result, a 2D image at each time point is visualized as a dot.
To covey the time information, they connected dots of two consecutive time points and colored them according to time.
Several researchers also used a similar approach to provide a visual summary of dynamic network data~\cite{von2015mobilitygraphs,van2016reducing, fujiwara2017visual}. 
On the other hand, J{\"a}ckle et al.~\cite{jackle2016temporal} visualized an MDS result in a 1D axis and used another axis to represent time. 
Since MDS may produce unnecessary rotation in the result, they reduced the rotations by flipping the $y$-coordinates based on their positions in the previous time point. 
Muelder et al.~\cite{muelder2016visual} also took a similar approach but they used a graph layout algorithm as a DR method instead of MDS.
However, all the approaches above have several problems. 
For example, when two modes in each matrix slice have different types (e.g., instances and variables), the DR result might not capture any useful patterns because each mode is mixed together when computing dissimilarities (refer to \autoref{sec:qual_comp} for concrete examples). 
Also, because each dot in the DR result represents a matrix, it is difficult to identify which instances or variables highly relate to a certain pattern appeared in the DR result (e.g., clusters) and, consequently, the result has low interpretability. 

While the visualizations above focus on showing the time points' similarities, some works are to overview the instance similarities over time. 
For example, Fujiwara et al.~\cite{fujiwara2018visual} used time-series distance measures, such as dynamic-time warping, to obtain the similarity of each instance's changes in a variable value across time.
Afterward, for each variable, they applied MDS or t-SNE to the computed similarities and then juxtaposed the DR results for different variables.  
Kesavan et al.~\cite{kesavan2020visual} extended the same approach for streaming high-dimensional data. 
In contrast to our framework, these approaches handle only one variable in each DR result. 

Recently, visualization researchers have started to use tensor decompositions~\cite{kolda2009tensor,lu2011survey} to analyze or simplify third-order tensors. 
The two most popular tensor decompositions are canonical decomposition (or CP decomposition)~\cite{harshman1970foundations,carroll1970analysis} and the Tucker decomposition~\cite{tucker1966some}.
CP decomposition expresses a tensor as the sum of a finite number of rank-one tensors (i.e., tensor-to-vector decomposition).
On the other hand, the Tucker decomposition can be considered as a high-order version of PCA and decomposes a tensor to a core tensor and a matrix along each mode (i.e., tensor-to-tensor decomposition).
For example, TPFlow~\cite{liu2018tpflow} introduces a similar method to CP decomposition, which finds the best slice of a space-time cube, where some meaningful patterns likely exist. 
Voila~\cite{cao2017voila} uses the Tucker decomposition to detect anomalies from a space-time cube. 
Also, TTHRESH~\cite{ballester2019tthresh} utilizes the Tucker decomposition to compress volume data into a smaller file size.
While these works extract important features or elements from third-order tensors, our framework generates an overview from a third-order tensor for visual identification of patterns, such as clusters and outliers, and provides interpretability in the DR result.

\begin{figure}[tb]
	\centering
	\captionsetup{farskip=0pt}
    \includegraphics[width=1.0\linewidth]{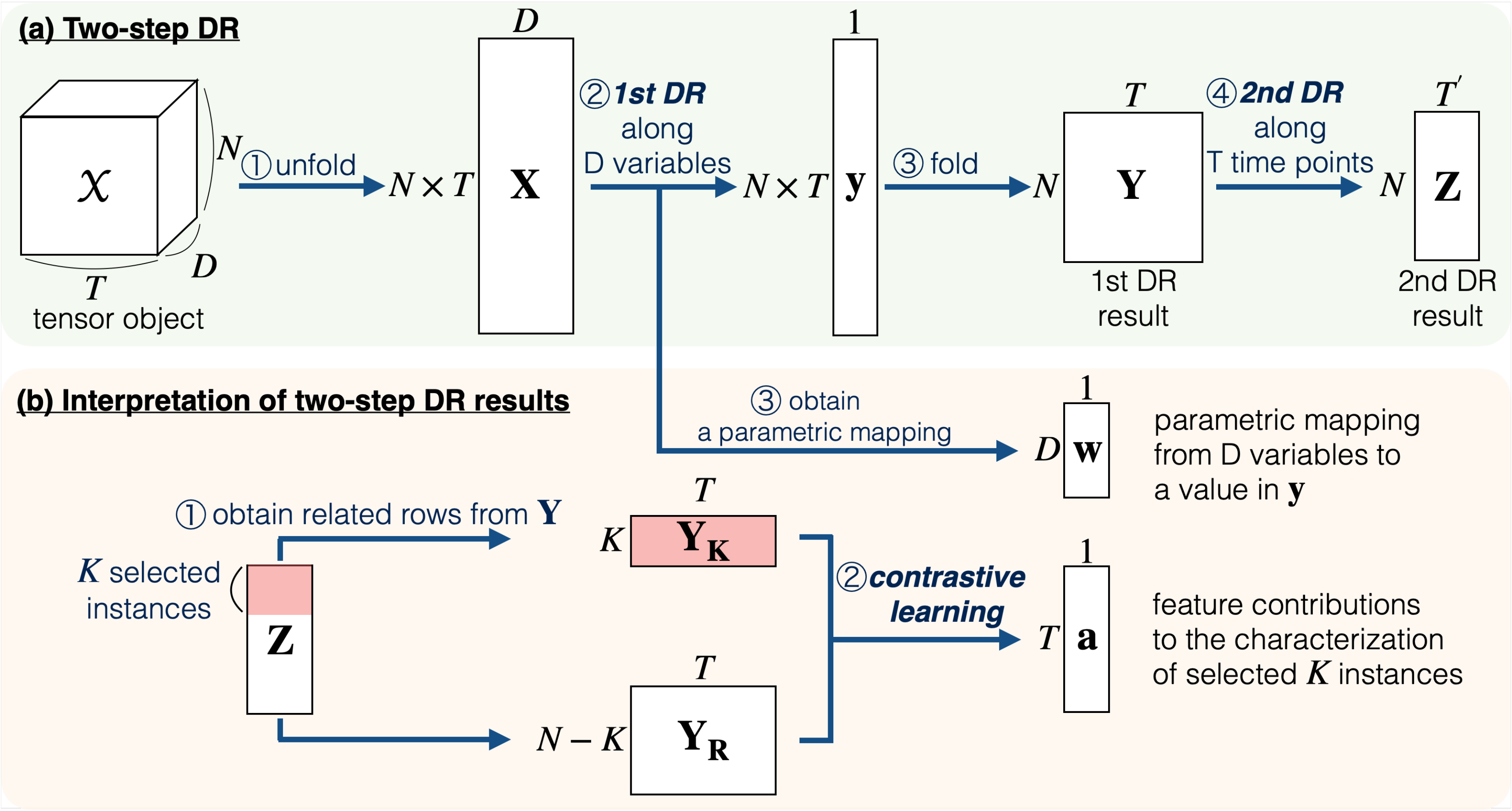}
    \caption{General architecture of \name{} back-end. Here, we demonstrate a case when showing instance similarities based on their temporal changes of variable values.}
	\label{fig:architecture}
\end{figure}

\section{Algorithm Architecture}
\label{sec:architecture}
\autoref{fig:architecture} shows a general architecture of the back-end of \name{}. \name{} provides two major functionalities: (a) two-step DR to project a third-order tensor onto a low-dimensional space and (b) generation of essential information for interpreting the two-step DR results. 

\subsection{Two-Step DR}
\label{sec:two_step_dr}
We describe how \name{} achieves the projection of a third-order tensor. 
To make an explanation concrete and concise, we use a case shown in \autoref{fig:architecture}a. 
The descriptions below related to $T$ time points, $N$ instances, and $D$ variables are interchangeable between themselves. 

The first step of DR is to compress a third-order tensor \Tensor{x} $\in \mathbb{R}^{T \times N \times D}$ into a matrix \FirstDR{} $\in \mathbb{R}^{N \times T}$, where certain information (e.g., variances) of variables is preserved as much as possible. 
To achieve this, we first apply \textit{tensor unfolding}~\cite{kolda2009tensor} along a variable mode (\autoref{fig:architecture}-a\textcircled{\small 1}), which reshapes \Tensor{X} to a matrix \Mat{X} of ($N \times T$) rows and $D$ columns by arranging all vectors (or often called \textit{fibers}) of $D$ length obtained through the slicing of \Tensor{X} along both time and instance modes. 
Afterward, we apply a DR method to \Mat{X} and reduce $D$ dimensions to 1 dimension (\autoref{fig:architecture}-a\textcircled{\small 2}).
Note that the similar approach is used in Unfold PCA~\cite{kiers1991hierarchical,kiers2000towards} to produce a matrix of ($N \times T$) rows and $D'$ columns where $D' < D$ and $D'$ is typically two or more as its purpose is obtaining the DR result with one step, unlike our two-step DR.
Now, \Tensor{X} is compressed into a vector \Vec{y} of length $(N \times T)$. 
Based on \Vec{y}'s indexes correspond to the time and instance directions of \Tensor{X}, we can fold \Vec{y} into a matrix \FirstDR{} $\in \mathbb{R}^{N \times T}$ (\autoref{fig:architecture}-a\textcircled{\small 3}).
Because the main purpose of the first DR is preserving the information of variables, a linear DR method that can be used for data compression is suitable.
For example, while we can use PCA to preserve the variances of variables, linear discriminant analysis (LDA) is also a potential option if the analyst wants to preserve differences between \Tensor{X} and another third-order tensor. 
Also, the linearity of DR is important to provide interpretability, as described in \autoref{sec:support_interpretability}.

The second step of DR is to visualize \FirstDR{} in a lower-dimensional space.
For this step (\autoref{fig:architecture}-a\textcircled{\small 4}), based on the analysis purpose, we can simply select any DR method that can be applied to a matrix, such as PCA and t-SNE~\cite{maaten2008visualizing}.
Through this step, \FirstDR{} $\in \mathbb{R}^{N \times T}$ can be represented as \SecondDR{} $\in \mathbb{R}^{N \times T'}$ ($T' < T$, typically $T' \in \{1, 2, 3\}$).

Instead of using the two-step DR above, similar to Unfold PCA~\cite{kiers1991hierarchical,kiers2000towards}, another potential approach is unfolding \Tensor{X} to a matrix of $N$ rows and ($D \times T$) columns and then apply DR in order to reduce dimensions of ($D \times T$) to a lower number of dimensions. 
When compared with this approach, the two-step DR has the main advantage in handling different modes (e.g., variables and time points) with clear distinction, and this benefits both identification and interpretation of patterns in \Tensor{X}. 
For example, when the analyst wants to review the similarities of $N$ instances mainly based on patterns seen along a time mode but not a variable mode, they can use the process shown in \autoref{fig:architecture}a. 
Also, as described in the next subsection, the interpretation of the DR results becomes more straightforward because, for example, we can understand which time points highly contribute to the characteristics of a cluster seen in the DR result.  
We provide more detailed comparisons in \autoref{sec:qual_comp}.

There are six different combinations to generate the two-step DR result \SecondDR{} based on which modes are selected as the first and second DR targets: (1\textsuperscript{st} DR target mode, 2\textsuperscript{nd} DR target mode) = \{(time, instance), (time, variable), (instance, time), (instance, variable), (variable, time), (variable, instance)\}. 
The analyst can choose a preferable combination from these based on their analysis interest. 
For example, when selecting (variable, time), a two-step DR result shows instance similarities mainly based on temporal behaviors while considering distribution differences in variables. 
On the other hand, a selection of (variable, instance) generates time points' similarities based on instances' states (i.e., values of the compressed variables each instance has) at each time point.

\subsection{Supporting Interpretability}
\label{sec:support_interpretability}

When analyzing the DR result, we often want to identify clusters from the DR result and understand the characteristics of the clusters~\cite{brehmer2014visualizing,nonato2018multidimensional,fujiwara2019supporting}.
Similar to the existing DR methods, identification of clusters can be visually performed on the two-step DR result (i.e., finding a set of points placed closely to each other). 
However, when compared with the case of applying ordinary DR methods to a matrix, understanding the cluster's characteristics from a two-step DR result is more complicated. 
Therefore, we provide algorithmic support for this task. 

As shown in \autoref{fig:architecture}b, \name{} provides two different pieces of information for the interpretability: \FC{}, feature contributions of $T$ time points to a cluster's characteristics, and \ParaMap{}, a parametric mapping used to compress $D$ variables into one dimension.

To obtain feature contributions \Vec{a}, we follow the contrastive learning (CL) based approach introduced by Fujiwara et al.~\cite{fujiwara2019supporting}.
As shown in \autoref{fig:architecture}-b\textcircled{\small 1}, \name{} first takes $K$ instance indices related to a target cluster and then, from \FirstDR{}, extracts \Mat{Y_K} $\in \mathbb{R}^{K \times T}$, a submatrix corresponding to these $K$ instances, and \Mat{Y_R}  $\in \mathbb{R}^{(N - K) \times T}$, the rest of \FirstDR{} (i.e., \Mat{Y_R} $=$ \FirstDR{} $\setminus$ \Mat{Y_K}). 
From inputs \Mat{Y_K} and \Mat{Y_R}, CL generates \FC{} $\in \mathbb{R}^{T}$ (\autoref{fig:architecture}-b\textcircled{\small 2}), which shows how strongly each time point contributes to the uniqueness of a target cluster with respect to the others.
By referring to \FC{}, the analyst knows which time points in \FirstDR{} they should review to understand the target cluster's characteristics.

However, each cell of \FirstDR{} represents the compressed variable from $D$ to 1 dimension. 
To understand the cluster's characteristics, we also need to know how the compressed variable is derived from the original $D$ variables. 
To do so, we can refer to a parametric mapping vector \ParaMap{} $\in \mathbb{R}^{D}$ (\autoref{fig:architecture}-b\textcircled{\small 3}), which is usually provided by DR methods for data compression (e.g., PCA). 
\ParaMap{} consists of a weight for each of $D$ variables, which is used to project $D$ variable values to one compressed value. 

\subsection{Implementation Example}
As described above, the back-end architecture of \name{} provides flexibility in the selection of the first DR, second DR, and CL. 
This flexibility enables \name{} to support various analysis needs, as discussed in \autoref{sec:two_step_dr}.
Here, we describe a representative implementation example, which we use through the rest of the paper. 
For the first DR, we use PCA because it is most popularly used for data compression when applying machine learning methods, including DR methods. 
We use UMAP~\cite{mcinnes2018umap} as the second DR because of its effectiveness to find patterns from nonlinear relationships. 
Also, unlike the other nonlinear DR methods (e.g., t-SNE~\cite{maaten2008visualizing}), UMAP is suitable for capturing both local and global topological structures of the data~\cite{mcinnes2018umap}. 
Because of this ability, UMAP is effective in finding patterns from both small- and large-scale data while many other nonlinear DR methods are not suitable for small-scale data (e.g., data with 50 instances). 
Lastly, for the purpose of understanding the characteristics of clusters, currently, ccPCA~\cite{fujiwara2019supporting} is the only available option; thus we use it as a CL method.

\begin{figure*}[tb]
	\centering
	\captionsetup{farskip=0pt}
    \includegraphics[width=0.95\linewidth]{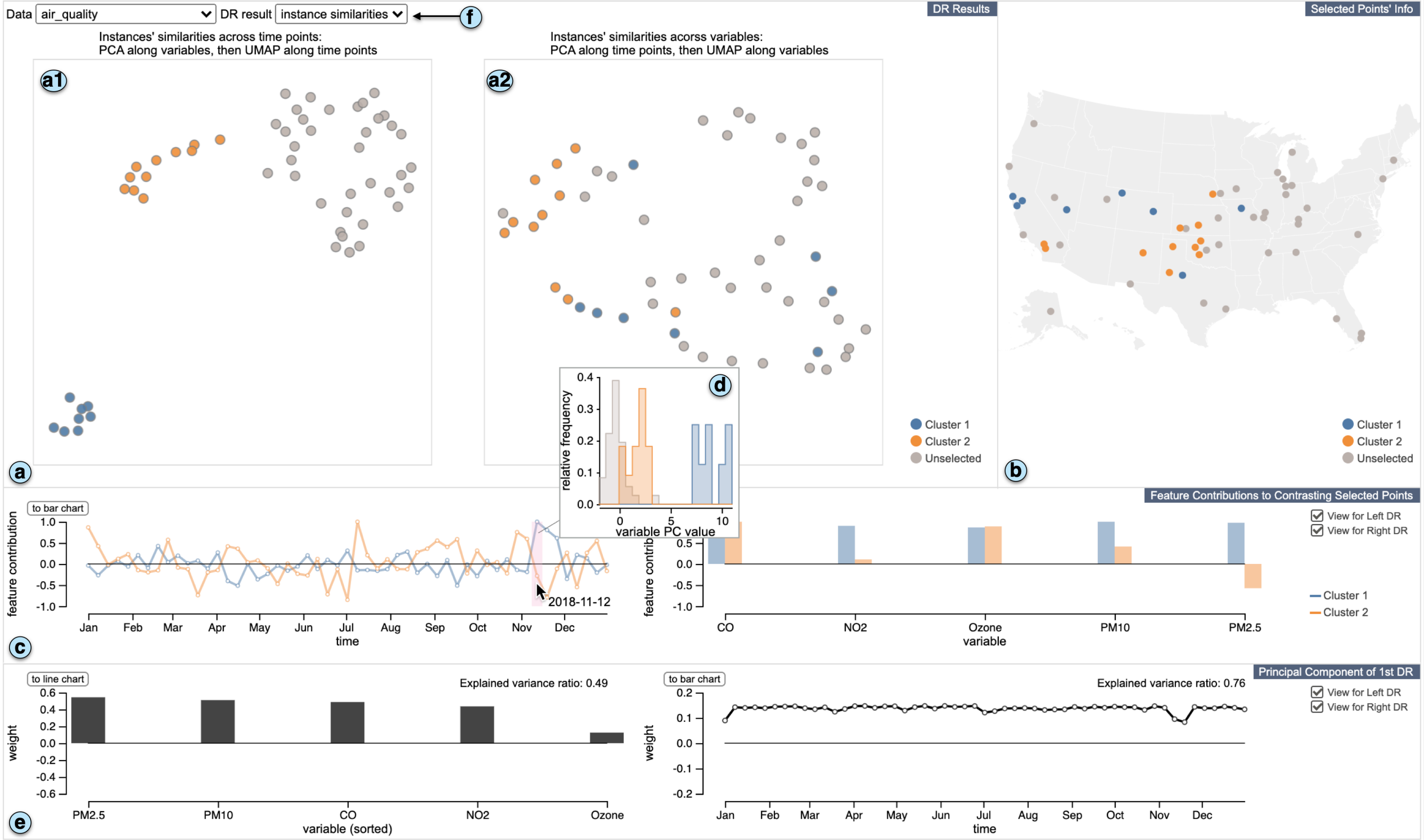}
    \caption{A screenshot of \name{} visual interface. 
    Here we visualize the AirData~\cite{airdata}, air quality data at outdoor monitors across the US, collected in 2018.
    (a) A two-step DR (TDR) view draws the DR results obtained through the two-step DR. 
    (b) A supplemental information (SI) view supports understanding selected points in the TDR view with the auxiliary information.
    (c) A feature contribution (FC) view visualizes features (either instances, variables, or time points) and their contributions to characteristics of each of the selected clusters.
    (d) A histogram comparison (HC) view shows the feature values in the first DR result \Mat{Y} of the selected element in (c). 
    (e) A parametric mapping (PM) view depicts parametric mappings generated in the first DR, specifically the mappings to the first principal component in this example. 
    (f) The analyst can select a type of DR results.}
	\label{fig:system_overview}
\end{figure*}

\begin{figure}[tb]
    \centering
    \includegraphics[width=1.0\linewidth]{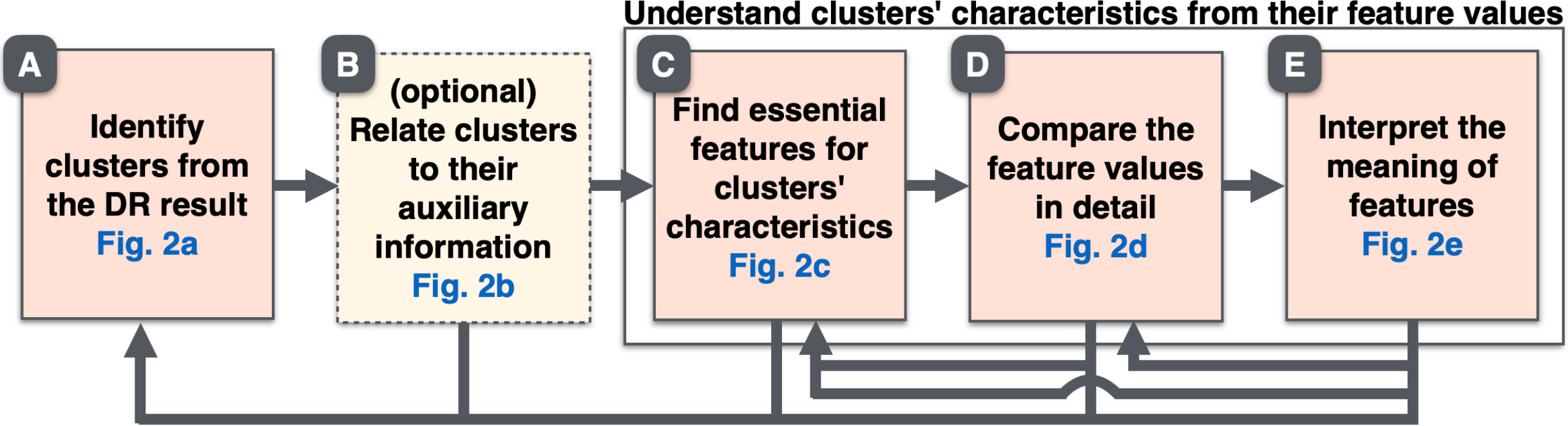}
    \caption{Multivariate time-series analysis workflow with \name{} visual interface.}
    \label{fig:workflow}
\end{figure}

\section{\name{} Visual Interface}
\label{sec:visual_interface}

\name{} provides a visual interface to support interactive analysis of the two-step DR results together with the information that helps the interpretation of the results. 
As shown in \autoref{fig:system_overview}, \name{} visual interface consists of five coordinated views: (a) a two-step DR (TDR) view, (b) a supplemental information (SI) view, (c) a feature contribution (FC) view, (d) a histogram comparison (HC) view, and (e) a projection mapping (PM) view.

\autoref{fig:workflow} shows an analysis workflow with \name{} visual interface. 
Here, we extend the workflow for high-dimensional data analysis introduced by Fujiwara et al.~\cite{fujiwara2019supporting} for multivariate time-series analysis. 
After obtaining a two-step DR result \SecondDR{}, feature contributions \FC{}, and a parametric mapping \ParaMap{}, the two-step DR result is visualized in \autoref{fig:system_overview}-a. 
The analyst can first visually identify clusters from the DR result (\autoref{fig:workflow}-A) and then analyze each cluster. 
When points in the DR result have the auxiliary information (e.g., the location information of instances), the analyst can (B) relate the identified clusters to such information, as shown in \autoref{fig:system_overview}-b. 
Afterward, they can move forward to the remaining steps (C, D, E), where the information of feature contributions and parametric mapping is used to understand the clusters' characteristics. 
With \autoref{fig:system_overview}-c, the analyst can start with (C) finding which features (i.e., columns in \FirstDR{}) highly contribute to characterizing each cluster.
For each of the highly contributed features, by using \autoref{fig:system_overview}-d, the analyst can (D) compare the differences of feature value distributions among clusters. 
Since the features are obtained through the compression with the first DR, the analyst also (E) interprets the meaning of the features by reviewing the parametric mapping information provided in \autoref{fig:system_overview}-e.
As indicated with the arrows in \autoref{fig:workflow}, the above steps often drive a continuous analysis loop in order to identify other clusters, select other features of interest, or examine findings obtained in the other view. 

For the rest of section, we describe each view of \name{} with a concrete analysis example using the US weekly air quality data in 2018~\cite{airdata}, which consists of 53 weeks, 55 counties, and 5 different air quality measures (i.e., $T=53$, $N=55$, $D=5$). 
A demonstration video of the interface is available at our online site~\cite{supp}.

\subsection{Visualization of Two-Step DR Results}
\label{sec:tdr_view}
The \TDRView{} (\autoref{fig:system_overview}-a) visualizes the results obtained through the two-step DR as scatterplots. 
As described in \autoref{sec:two_step_dr}, the two-step DR can generate six different results from a multivariate time-series dataset based on target modes of the first and second DR (e.g., the first DR along a variable mode and the second DR along a time mode). 
From these results, the analyst can select which mode's similarities they want to show from a drop-down menu at \autoref{fig:system_overview}-f. 
For example, in \autoref{fig:system_overview}-a, instance similarities are selected.
Consequently, as described at the top of each of scatterplots (a1 and a2), the \TDRView{} shows two results that are obtained by applying the first and second DR along (variable, time) and (time, variable) at the left and right, respectively.

From the results, the analyst can visually identify clusters and manually select them by using a lasso selection. 
Selected points are labeled as one cluster and color-coded with a categorical color. 
For example, in \autoref{fig:system_overview}-a1, the analyst has first selected Cluster 1 (blue) and then Cluster 2 (orange).
In addition to the lasso selection, \TDRView{} also supports fundamental interactions, such as zooming and panning.
After the selection, all other views update their visualizations. 
From \autoref{fig:system_overview}-a1 and a2, we can see that although the orange points in a1 are also relatively placed closely to each other in a2, the clusters in a1 tend to be more mixed with each other in a2.
This indicates that instances (i.e., counties) in these clusters generally have similar temporal patterns in their representative variable values (i.e., representative air quality measure); however, they tend to have different patterns in variable values (e.g., some of the blue points may have high variable values but the others do not).
We demonstrate an analysis example utilizing both of the two different DR results in \autoref{sec:study_contact_networks}. 

\subsection{Visualization of Related Contexts}
After identifying clusters, we often want to understand what kind of points are included in each cluster and why they are clustered by the two-step DR. 

The \SIView{} (\autoref{fig:system_overview}-b) is designed for the former task. 
The \SIView{} visualizes the auxiliary information of the selected points in the \TDRView{} if available. 
In \autoref{fig:system_overview}-b, the location information of the selected counties is visualized, where the blue and orange clusters tend to be seen in more west and center, respectively. 
\name{} provides a set of predefined visualizations and selects one from them based on which mode and dataset need to be visualized. 
For example, when showing the information for a time mode of the air quality data, \name{} shows a calendar-based visualization to convey the seasonal patterns.
While the \SIView{} shows the location information for an instance mode of a geospatial dataset, when analyzing a network data, the \SIView{} can provide a node-link diagram. 
We demonstrate examples in \autoref{sec:cs}. 

\subsection{Visualization of Feature Contributions and Values}
In the next step, the analyst can analyze the clusters' characteristics with the \FCView{} (\autoref{fig:system_overview}-c) and \HCView{} (\autoref{fig:system_overview}-d).

The \FCView{} shows feature contributions \FC{} for each of the DR results in the \TDRView{} (the left and right plots in \autoref{fig:system_overview}-c correspond to \autoref{fig:system_overview}-a1 and a2, respectively).
In default, line charts are employed for feature contributions of time points, while bar charts are used for those of instances or variables. 
However, the analyst can switch line and bar charts by clicking the button placed at the top of each of $y$-axes (e.g., ``to bar chart'' at the left side of \autoref{fig:system_overview}-c).
Also, with the checkboxes placed at the far right in \autoref{fig:system_overview}-c, the analyst can select showing only one of the plots to use more screen space.
Since we obtain feature contributions for each cluster, we visualize them with the corresponding cluster color. 
\name{} scales feature contributions between $[-1, 1]$ by dividing each set of feature contributions by their maximum absolute value.
Closer to either $1$ or $-1$ indicates higher contributions to the characterization of a cluster. 
The meaning of the sign is discussed in \autoref{sec:sign_adj}.
For features that have high contributions, each cluster likely has different distributions from the other points. 

To compare value distributions of the selected feature, as shown in \autoref{fig:system_overview}-d, the \HCView{} shows relative frequency histograms of selected clusters (e.g., blue and orange) and unselected points (gray) with the corresponding colors. 
The $x$-axis of the histograms represents feature values (i.e., cell values in \FirstDR{}).
The $y$-axis shows relative frequency---a ratio of the number of items in each bin to the total number of items across all bins---within each group and its maximum limit is set to the maximum relative frequency among the histograms.
From the result in \autoref{fig:system_overview}-d, at the selected week highlighted with pink (i.e., a week of November-12th, 2018), the blue cluster tends to have much higher feature values than the others. 

\subsubsection{Sign Adjustment of Feature Contributions}
\label{sec:sign_adj}
\begin{figure}[tb]
	\centering
	\captionsetup{farskip=0pt}
    \includegraphics[width=0.995\linewidth]{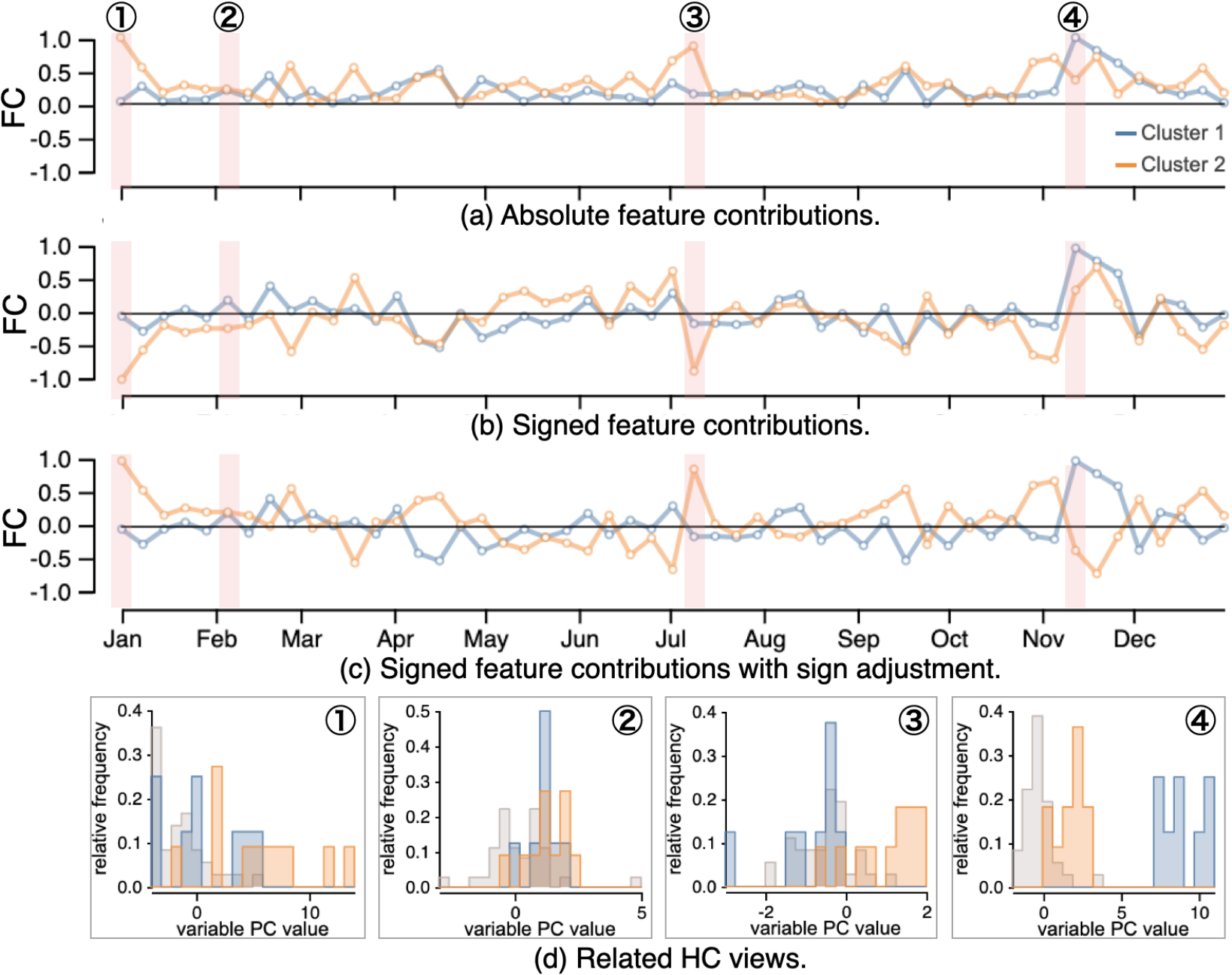}
    \caption{Comparison of visualizations of feature contributions (FCs): (a) absolute FCs, (b) signed FCs, and  (c) signed FCs with sign adjustment. (d) shows the \HCView{}s corresponding to the four selected time points \textcircled{\small 1}--\textcircled{\small 4} in (a), (b), and (c).}
	\label{fig:fc}
\end{figure}

ccPCA~\cite{fujiwara2019supporting}, which is used as a default CL method in \name{}, produces signed feature contributions (FCs). 
Signed FCs have a strength of differentiating features of having lower and having higher values within a selected cluster. 
For example, when looking at Cluster 2 (orange) in \autoref{fig:fc}-a, where the absolute FCs are shown, both time points \textcircled{\small 3} and \textcircled{\small 4} have relatively high FCs; however, as shown in \autoref{fig:fc}-d, while Cluster 2 tends to have high values at \textcircled{\small 3}, it has low values at \textcircled{\small 4}. 
On the other hand, the signed FCs shown in \autoref{fig:fc}-b indicates the difference of time points \textcircled{\small 3} and \textcircled{\small 4} (\textcircled{\small 3}: negative sign, \textcircled{\small 4}: positive sign). 

Despite the usefulness of signed FCs, similar to ordinary PCA, the \textit{sign ambiguity} problem~\cite{turkay2017designing,fujiwara2019incremental,fujiwara2019supporting} in ccPCA limits their interpretability. That is, the signs are arbitrarily selected, and thus they do not reflect whether features contribute to having higher or lower values than others. 
For example, in \autoref{fig:fc}-b, although both Clusters 1 and 2 have similar line shapes and strong positive FCs at \textcircled{\small 4}, as shown in \autoref{fig:fc}-d, at this time point, Cluster 1 (blue) tends to have high values while Cluster 2 (orange) tends to have low values. 
Therefore, using the signed FCs directly produced by ccPCA might mislead the analyst (e.g., they might consider Clusters 1 and 2 are similar from \autoref{fig:fc}-b).

To solve the above problem, we introduce a sign adjustment algorithm that optimally matches the directions of sign and value distributions (i.e., when a sign is positive, a cluster tends to have higher feature values than others, and vice versa).
First, for each feature, we compute Pearson's correlation coefficient $r$ ($-1 \leq r \leq 1$) between all points' cluster memberships (i.e., 0: points are non-members, 1: points are members of the selected cluster) and their feature values.
When $r$ is closer to $1$, members of the cluster more likely have a higher feature value than non-members. 
On the other hand, closer to $-1$, higher possibility to have a lower feature value. 
We denote a set of $r$ for all features as a vector \Vec{r}. 
Next, we compute a score of agreement $s$ between correlation coefficients \Vec{r} and signed FCs \FC{} by taking their dot product (i.e., $s = $ \Vec{r} $\cdot$ \FC{}).
$s$ increases when an element of \Vec{r} and the corresponding element of \FC{} have the same signs, while $s$ decreases when they have the opposite signs. 
Also, the magnitudes of elements of \Vec{r} and \FC{} can be considered as weights to decide how much $s$ should increase or decrease.
As a result, $s$ becomes a higher positive value when each pair of elements of \Vec{r} and \FC{} has higher magnitudes of $r$ and FC with the same signs. 
When $s < 0$, \Vec{r} and \FC{} disagree with each other; thus, we flip signs of all elements in \FC{}.

The result after applying the sign adjustment is shown in \autoref{fig:fc}-c.
Now, we can see that Clusters 1 and 2 have clearly different patterns in FCs. 
For example, while Cluster 1 has a strong positive FC at \textcircled{\small 4}, Cluster 2 has strong positive FCs at \textcircled{\small 1} and \textcircled{\small 3} and a strong negative FC at \textcircled{\small 4}.
Also, by referring to the \HCView{}s in \autoref{fig:fc}-d, these differences well represent differences in the distributions of feature values. 
For instance, at \textcircled{\small 1}, Cluster 2 tends to have high feature values but low feature values at \textcircled{\small 4}, while Cluster 1 has high feature values at \textcircled{\small 4}. 
With the sign-adjusted FCs, to understand the differences between clusters, the analyst can mainly focus on reviewing features that have highly different contributions between clusters.

Note that the work that introduced ccPCA~\cite{fujiwara2019supporting} also presented a sign adjustment algorithm to deal with the inconsistency of signs of FCs across clusters.
However, our algorithm focuses on matching the directions of sign and value distributions for each cluster to ensure that a\,cluster\,has\,high\,feature\,values\,when\,its\,feature\,has\,a\,strong\,positive\,FC.

From the result shown in \autoref{fig:system_overview}-c(left), now we know Cluster 1 tends to have high feature values around the middle of November but low feature values around the middle of April.
Also, we can see that Cluster 2 tends to have the opposite patterns from Cluster 1. 

\subsection{Visualization of Parametric Mappings}
The last analysis step is to understand the meaning of features obtained after the first DR of the two-step DR (i.e., columns in \FirstDR{}).  
To support this task, the \PMView{} (\autoref{fig:system_overview}-e) visualizes a vector of parametric mapping \ParaMap{} for each of the DR results as either line or bar chart, as similar to the \FCView{}.
Note that \ParaMap{} is common across all points, and thus all lines or bars are colored in black.
Also, using texts, at the top-right corner of each plot in the \PMView{}, we inform the quality of the first DR (e.g., \textit{explained variance ratio} provided by many linear DR methods such as PCA and LDA).
From \autoref{fig:system_overview}-e(left), we can see that the feature values are generated with similar weights for all measures except for ``Ozone''. 
Therefore, we can interpret the feature values in \autoref{fig:system_overview}-d are close to the mean of ``PM2.5'', ``PM10'', ``CO'', and ``NO2''.

\subsection{Implementation}

We have developed \name{} as a web application\footnote{The source code is available at our online site\cite{supp}}.
For the back-end of \name{}, including the algorithms described in \autoref{sec:architecture}, the sign adjustment algorithm in \autoref{sec:sign_adj}, and the generation of histogram information for the \HCView{}, we use Python to integrate all the existing implementations, such as UMAP~\cite{mcinnes2018umap} and ccPCA~\cite{fujiwara2019supporting}.
The front-end visual interface is implemented with a combination of HTML5, JavaScript, D3~\cite{bostock2011d3}, and WebGL. 
We use WebSocket to communicate between the front-end and back-end.

\section{Case Studies}
\label{sec:cs}

We have shown the effectiveness of \name{} through the analysis of the air quality data~\cite{airdata}. 
Here we further analyze the same data from different aspects. 
Additionally, we demonstrate three additional case studies, including analyses of a body sensing dataset, a dynamic social network, and supercomputer's hardware logs.
For each study, we have preprocessed each dataset to deal with its missing values or extract useful information for the analysis. 
All the processed datasets (except for the supercomputer's hardware logs due to their confidentiality) and parameters used for each DR result are available at our online site~\cite{supp}.

\subsection{Study 1: Analysis of US Air Quality Data}
\label{sec:cs1}

\begin{figure}[tb]
    \centering
    \includegraphics[width=1.0\linewidth]{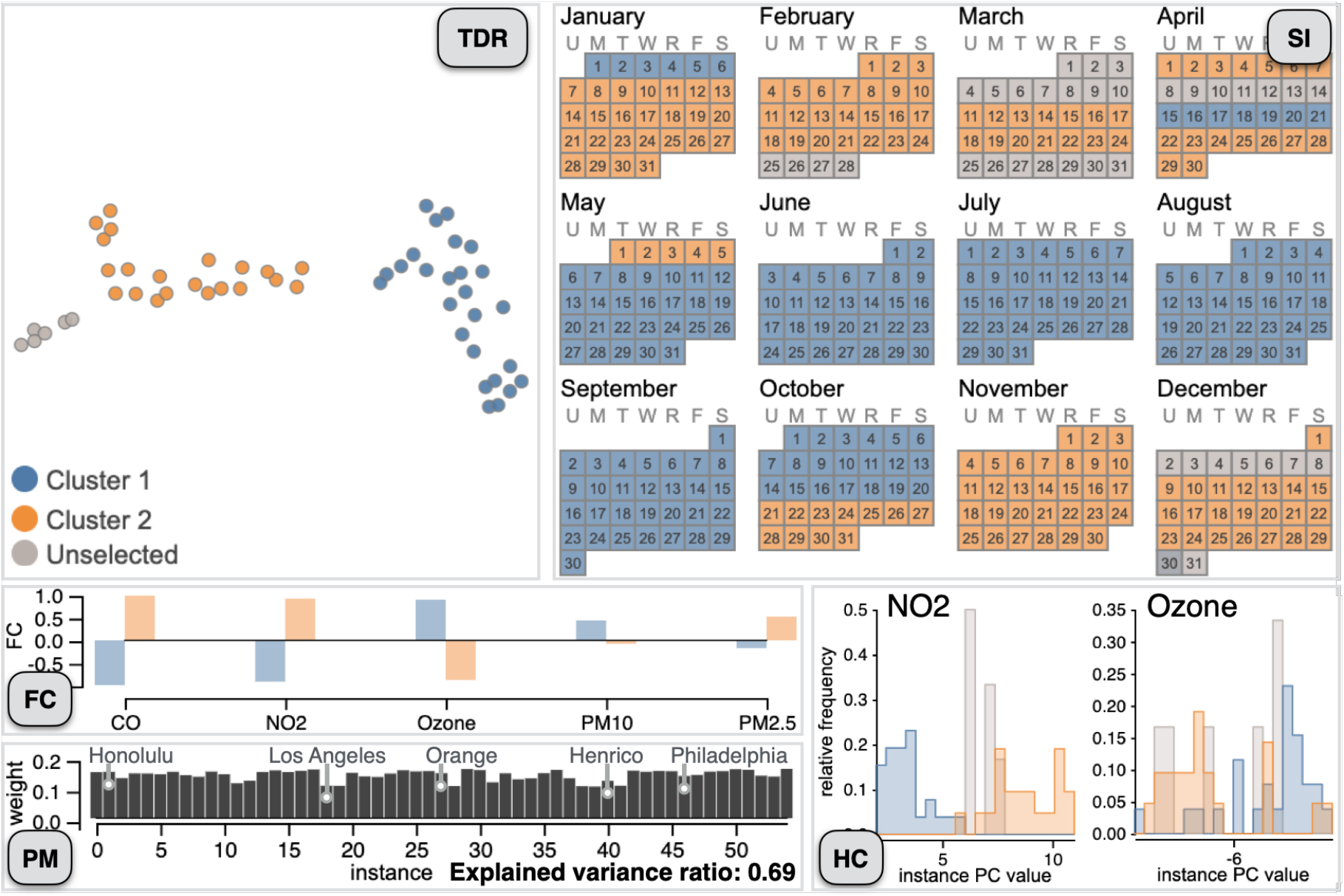}
    \caption{Case study 1. (TDR) shows similarity of each week's air quality measures. 
    (SI, FC, PM) are the SI, FC, \PMView{}s after selecting Clusters 1 and 2 in the \TDRView{}, respectively. 
    (HC) shows the \HCView{}s when selecting ``NO2'' and ``Ozone'' form (FC).}
    \label{fig:cs1_2}
\end{figure}

\vspace{2pt}
\noindent\textbf{Analysis of Weekly Patterns of Air Quality Measures.}
In \autoref{sec:visual_interface}, we have analyzed the clusters of instances (i.e., US counties) selected in \autoref{fig:system_overview}-a1;
here, we analyze the similarities of time points (weeks in 2018) based on their values of air quality measures. 
For this task, we apply the two-step DR using PCA along an instance mode (i.e., counties) and then UMAP along a variable mode (i.e., air quality measures).
The generated results are shown in \autoref{fig:cs1_2}.

From the \TDRView{} shown in \autoref{fig:cs1_2}-TDR, we select several clearly separated points (i.e., weeks) as clusters. 
For this data, the \SIView{} provides a calendar-based visualization and indicates the corresponding weeks for each cluster (\autoref{fig:cs1_2}-SI). 
We notice that while the blue cluster generally relates to the weeks from May to the middle of October, the orange cluster consists of the weeks from the late fall to the early spring.
To understand the differences of each cluster, we refer to the \FCView{} in \autoref{fig:cs1_2}-FC.
The two clusters have quite different FCs, and thus seem to have different feature values as well.
For example, in the histograms of ``NO2'', as shown in \autoref{fig:cs1_2}-HC, Clusters 1 (blue) and 2 (orange) tend to have low and high feature values when compared with others, respectively.
On the other hand, in the histograms of ``Ozone'', we can see the opposite distributions.
From the \PMView{} (\autoref{fig:cs1_2}-PM), we can see that several counties (e.g., Honolulu) have slightly higher weights than others when generating the feature values.

In general, we can conclude that the air quality data has seasonal changes, such as ``NO2'' has higher values around the winter (orange weeks) when compared with around the summer (blue weeks).

\subsection{Study\,2:\,Analysis\,of\,MHEALTH\,(Mobile\,Health)\,Dataset}

In this case study, we analyze the MHEALTH (Mobile HEALTH)~\cite{mhhealth1,mhhealth2} dataset.
This dataset consists of physical recordings of motion and vital signs for ten volunteers while performing twelve physical activities. Sensors are placed on the subjects' chest, wrist, and ankle. The measurements taken from sensors include movement experienced by different body parts, such as acceleration with the magnitude for each of X-, Y- and Z-directions. 
The sensor modalities are recorded at a sampling rate of \SI{50}{\hertz}. 
The dataset contains points that represent bursts of highly active minutes.

\begin{figure}[tb]
    \centering
    \includegraphics[width=1.0\linewidth]{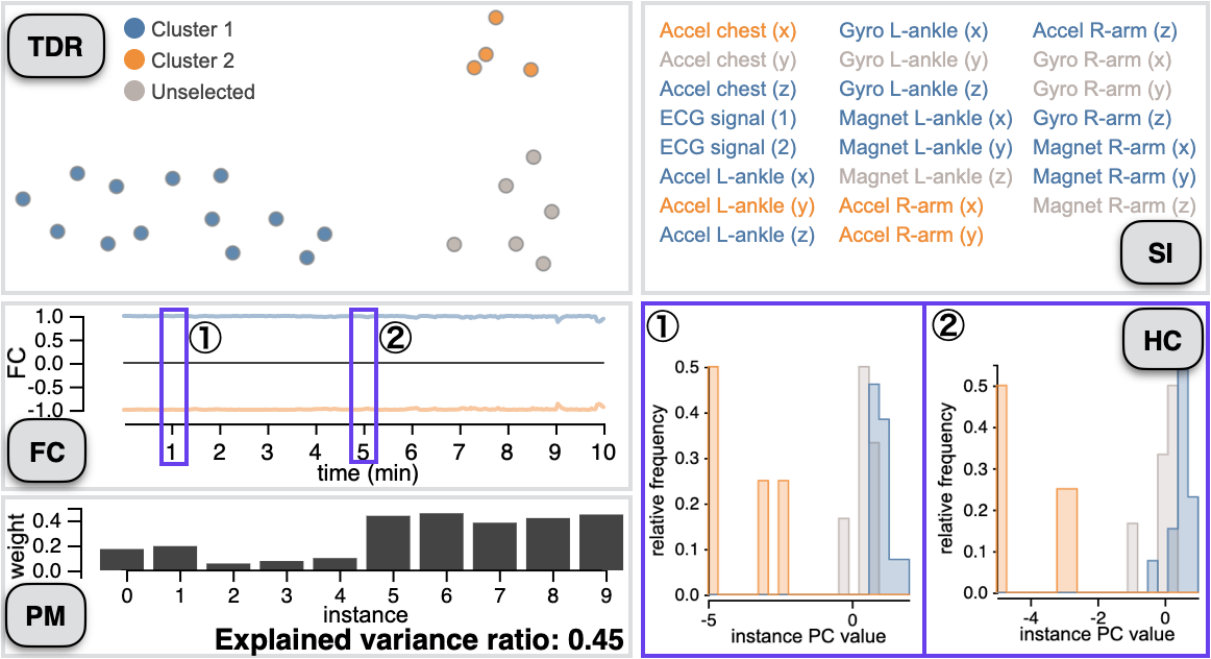}
    \caption{Case study 2-1. 
    (TDR) shows similarities of physical activity measurements based on their temporal behaviors. 
    (FC, PM) are the FC and \PMView{}s after selecting two clusters from (TDR). 
    (HC) shows the \HCView{}s after selecting two different timestamps from (FC).
    }
    \label{fig:cs2_1}
\end{figure}

\vspace{2pt}
\noindent\textbf{Study 2-1: Categorization of Physical Activity Measurements.} 
As shown in \autoref{fig:cs2_1}, from the \TDRView{} (\autoref{fig:cs2_1}-TDR), where variables' (i.e., measurements') similarities are shown by applying PCA and UMAP along instance and time modes, respectively, we select Clusters 1 and 2 (blue and orange).
The SI view in \autoref{fig:cs2_1}-SI lists all measurements related to each cluster as texts. 

Afterward, we review the related information with the FC, PM, and HC views (\autoref{fig:cs2_1}-FC, PM, HC).
From \autoref{fig:cs2_1}-FC, we can see that, across time, blue and orange clusters have strong positive and negative FCs, respectively. 
To further investigate, we select several timestamps and review the corresponding histograms.
\autoref{fig:cs2_1}-HC shows the histograms at two examples of the selected timestamps (\textcircled{\small 1} and \textcircled{\small 2}).
From the results shown in \autoref{fig:cs2_1}-HC, all the measures within each cluster tend to have close feature values (e.g., in \textcircled{\small 1}, all orange points have low feature values).
By looking at \autoref{fig:cs2_1}-PM, PCA seems to generate the feature values with higher weights for Subjects 5--9 when compared to Subjects 0--4 with an explained variance of $0.45$. 

From the observations above, we can say that all the measures in Cluster 2, which includes the accelerations of the chest (X-direction), the left-ankle (Y-direction), and the right-arm (X- and Y-directions) have similar value distributions for each timestamp. 
The same applies to the measures in Cluster 1.

\begin{figure}[tb]
    \centering
    \includegraphics[width=1.0\linewidth]{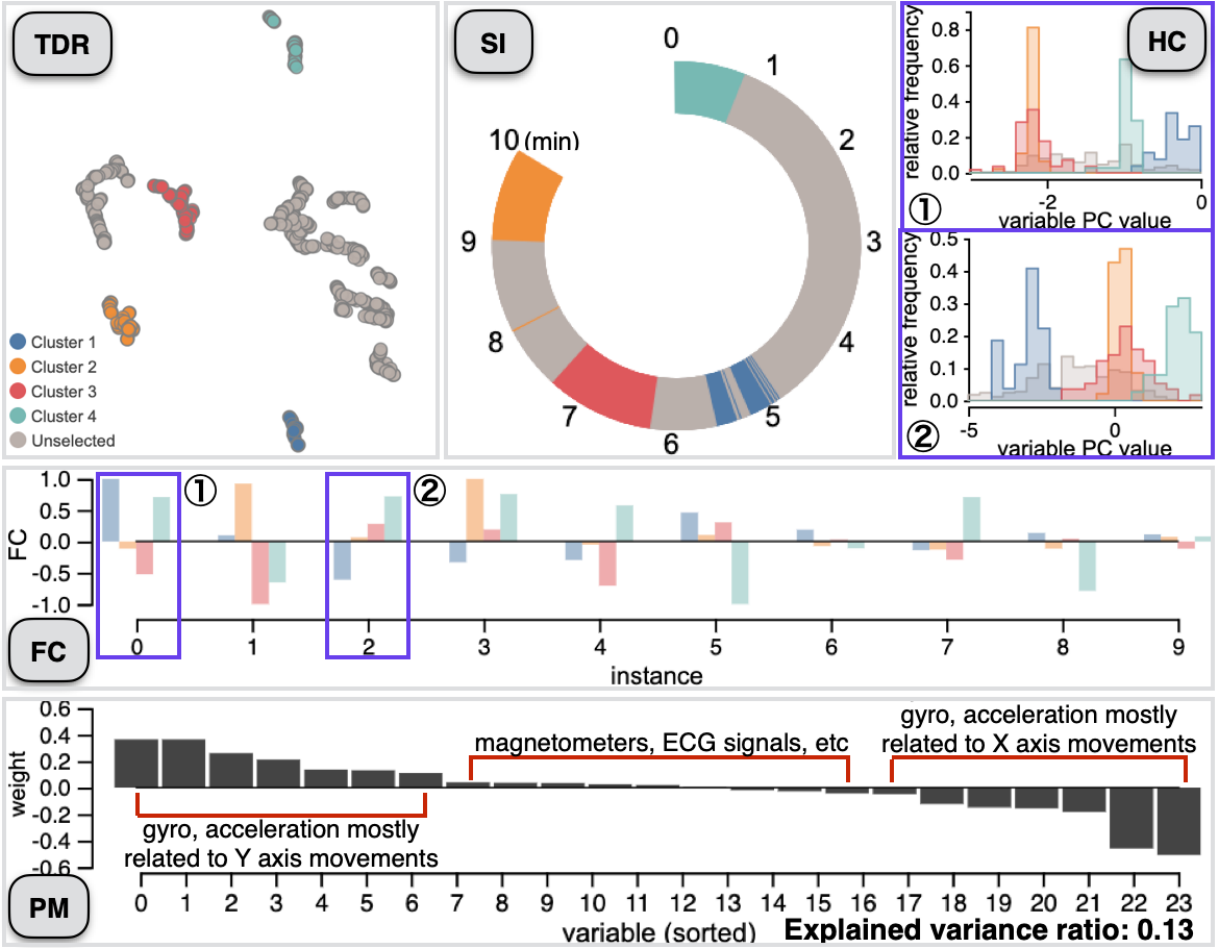}
    \caption{Case study 2-2. 
    (TDR) shows similarities of timestamps based on subjects' activities. 
    (SI) visualizes the corresponding time information with a circular layout.
    (FC, PM) are the FC and \PMView{}s after selecting four clusters from (TDR). 
    (HC) shows the \HCView{}s after selecting two different instances from (FC).}
    \label{fig:cs4_1_t}
\end{figure}

\vspace{2pt}
\noindent\textbf{Study 2-2: Classification of Temporal Patterns among Subjects.}
Next, we analyze the similarities of time points in the duration of activity measurement (10 minutes).
During the measurement, the subjects performed an activity set, including standing still, walking, running, etc. 
We apply the two-step DR using PCA along a variable mode and then UMAP along an instance mode.
The generated results are shown in \autoref{fig:cs4_1_t}.
From the \TDRView{} shown in \autoref{fig:cs4_1_t}-TDR, we select multiple clearly separated clusters. 
In the SI view (\autoref{fig:cs4_1_t}-SI), we see that each cluster is gathered together with a range of about 1 minute. 
Since the subjects were asked to perform each activity with a duration of approximately 1 minute for collecting data, we can expect that each cluster well represents each of the activities. 

To understand the differences of each cluster, we refer to the other views (\autoref{fig:cs4_1_t}-FC, HC, and PM).
All clusters have quite different FCs in \autoref{fig:cs4_1_t}-FC. 
Also, by referring to the \HCView{}s, as the examples in \autoref{fig:cs4_1_t}-HC show, each subject tends to have quite different feature values.
For example, while Subject 0 (annotated with \textcircled{\small 1}) tends to have high feature values during the activity corresponding to Cluster 1 (blue), Subject 2 (annotated with \textcircled{\small 2}) tends to have low feature values for Cluster 1.
From the \PMView{}, we understand that the feature values mainly relate to the measures of gyroscopes and accelerations but not magnetometers or ECG signals. 
More specifically, the measures related to Y-directions (e.g., ``gyro R-forearm (y)'', ``gyro L-ankle (y)'', ``acceleration chest (y)'') tend to have positive weights, while the measures of X-direction have negative weights. 

Therefore, we can conclude that, in general, the two-step DR successfully separates time points related to a specific activity based on the differences of each activity in the measures of X- and Y-directions.

\begin{figure}[tb]
    \centering
    \includegraphics[width=1.0\linewidth]{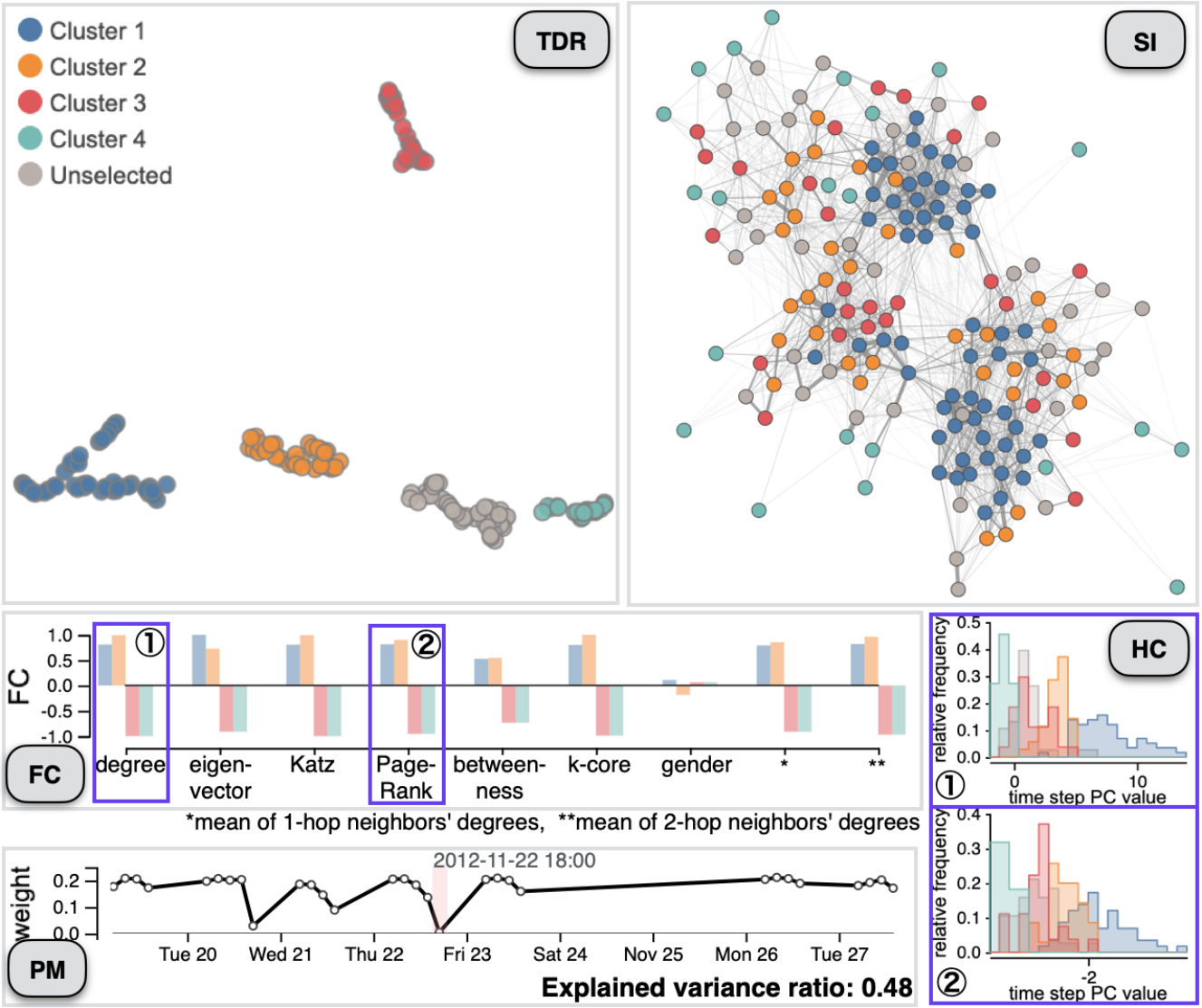}
    \caption{Case study 3-1. (TDR) shows similarities of the students based on their node features obtained with DeepGL~\cite{rossi2018deep}. (SI) draws a node-link diagram of the entire contact network. (FC, PM) are the FC and PM views after selecting four clusters from (TDR). 
    (HC) shows the HC views after selecting two different node features from (FC).}
    \label{fig:nw_1}
\end{figure}

\begin{figure}[tb]
    \centering
    \includegraphics[width=1.0\linewidth]{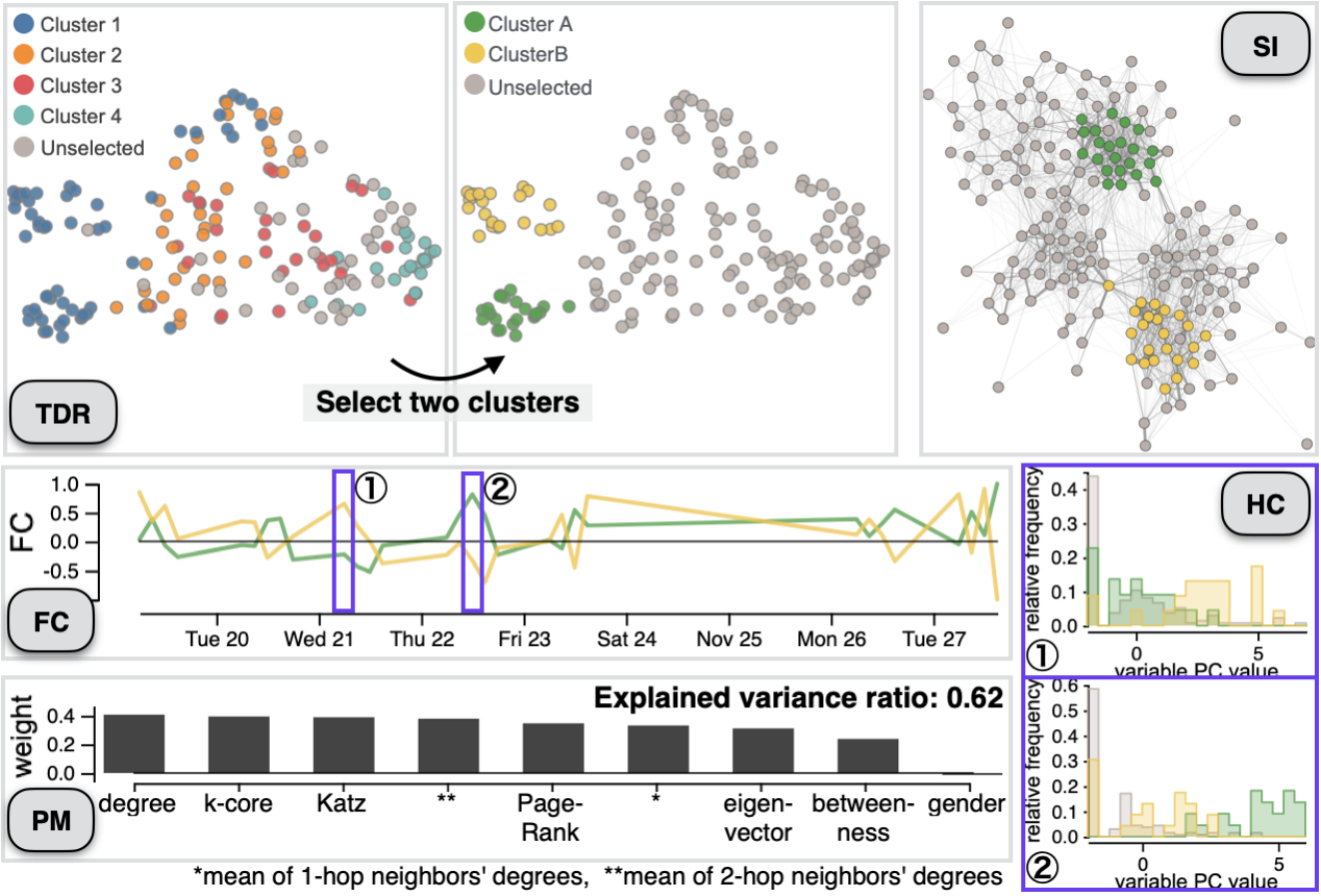}
    \caption{Case study 3-2. (TDR) shows similarities of the students based on their temporal behaviors. From the result at the left, where the colors correspond to the selection in \autoref{fig:nw_1}-TDR, we select Clusters A and B at the right. 
    (SI, FC, PM) are the SI, FC, and PM views after selecting the two clusters.
    (HC) shows the HC views after selecting two different timestamps from (FC).}
    \label{fig:nw_2}
\end{figure}

\subsection{Study 3: Analysis of Dynamic Contact Networks}
\label{sec:study_contact_networks}
Here, we provide an analysis example of dynamic networks, using a dataset of contacts between high school students in Marseilles, France~\cite{fournet2014contact}.
This dataset contains network links, which represent the students' face-to-face contacts collected with 20-second intervals for several days.
We construct temporal snapshots from this dynamic network by aggregating contacts within a time window of 5--9~AM, 10~AM--12~PM, 1--3~PM, 4--6~PM, or after 6~PM for each day.
This procedure generates 30 snapshots (i.e., networks) of 180 students (i.e., nodes) with the mean of 193 contacts (i.e., links). 
To extract features of network nodes, we apply DeepGL~\cite{rossi2018deep}, a network representation learning method that produces features consisting of node attributes (e.g., gender), network centralities (e.g., degree centrality~\cite{newman2018networks}), network measures (e.g., $k$-core number~\cite{newman2018networks}), and those of statistical values of neighbors (e.g., the mean degree centrality of 1-hop neighbors). 
As a result, a tensor of $T=30$, $N=180$, and $D=10$ is generated. 

\vspace{2pt}
\noindent\textbf{Study 3-1: Categorization of Students Based on Node Features.}
\autoref{fig:nw_1}-TDR shows a result generated by the two-step DR using PCA along a time mode and UMAP along a variable mode (i.e., the dots in the TDR view represent instances).
The result contains five distinct clusters and we select four of them. 
The resultant visualizations are shown in \autoref{fig:nw_1}-SI, HC, FC, and PM. 
Here, the \SIView{} draws an overall network constructed using a time window of the entire measurement period. 
From \autoref{fig:nw_1}-SI, we notice that the blue nodes (i.e., students) in Cluster 1 can be seen in the strongly connected regions.
In contrast, the teal nodes in Cluster 4 have only a small number of links to the others.

To further understand each cluster's characteristics, we review the \FCView{} (\autoref{fig:nw_1}-FC). 
We can see that, except for ``gender'', generally Clusters 1 and 2 have strong positive FCs, while Clusters 3 and 4 have strong negative FCs. 
In the \HCView{}s (\autoref{fig:nw_1}-HC), which show the histograms of \textcircled{\small 1} degree and \textcircled{\small 2} PageRank, Clusters 1 and 2 tend to have higher values than Clusters 3 and 4.
Especially, Cluster 1 has much higher values than the others.
From the \PMView{} (\autoref{fig:nw_1}-PM), we can see that the features in the \FCView{} are generated by using large weights around mornings whereas evenings have close to zero weights (e.g., 6~PM on November 22nd, as highlighted). 
This indicates that the first DR (i.e., PCA) has effectively selected time points where the students more actively contacted each other. 

With the above analysis, we can conclude that during school hours, the students in Cluster 1 played a central role in communications among students as they have high values for various network centralities.

\vspace{2pt}
\noindent\textbf{Study 3-2: Categorization of Students Based on Temporal Communication Patterns.}
Together with the results in Study 3-1, we further review the instance similarities obtained by applying PCA along a variable mode and UMAP along a time mode. 
\autoref{fig:nw_2}-TDR(left) shows the two-step DR result colored based on the selection in \autoref{fig:nw_1}-TDR. 
We can see that each of the currently selected clusters is generally arranged from left to the right. 
However, most students in Cluster 1 are separated into two distinct clusters in the far left.
We select the two clusters as Clusters A (green) and B (yellow), as presented in \autoref{fig:nw_2}-TDR(right).
In the SI view (\autoref{fig:nw_2}-SI), these two clusters are clearly separated into the two strongly connected regions at the top and bottom. 
The FC view (\autoref{fig:nw_2}-FC) shows that the two clusters have different patterns in FCs across time. 
We select two clear peaks, \textcircled{\small 1} (Wednesday morning) and \textcircled{\small 2} (Thursday noon), to see their value distribution differences with the \HCView{} (\autoref{fig:nw_2}-HC).
We can see that Clusters A and B tend to have high feature values at \textcircled{\small 2} and \textcircled{\small 1}, respectively.
By looking at the \PMView{} (\autoref{fig:nw_2}-PM), the feature values represent the network centralities and measures but not gender.

From the above observations, we can say that the students in Cluster 1 (the central role of the communications) can be further categorized into two different groups, Clusters A and B, which had active communications at the different time periods.

\subsection{Study 4: Analysis of Supercomputer Hardware Logs}
In this study, we analyze hardware logs obtained from a supercomputer. 
Supercomputers are required to have high robustness and reliability to continuously run large-scale computations. 
Analyzing their hardware logs is fundamental to revealing and understanding abnormal hardware behaviors (e.g., extreme increases in CPU temperatures), which can lead to hardware failures or errors~\cite{guo2018valse,shilpika2019mela}. 

Here, we specifically review the K computer's~\cite{miyazaki2012overview} hardware logs on January-12th, 2017.
The logs were obtained from 864 compute racks, where 1,163 different measures (e.g., CPU temperatures, circuit voltages, and cooling fan
spin speeds) are collected every 5 minutes (i.e., 1,440 timestamps in a day).
Therefore, the logs can be represented as a $T \times N \times D$ tensor where $T = 1,440$, $N = 864$, and $D = 1,163$ (in total, more than 1.4 billion elements). 
Through this case study, we demonstrate how \name{} helps the analyst identify and characterize outliers from an extremely large-scale dataset.

\begin{figure}[tb]
    \centering
    \includegraphics[width=1.0\linewidth]{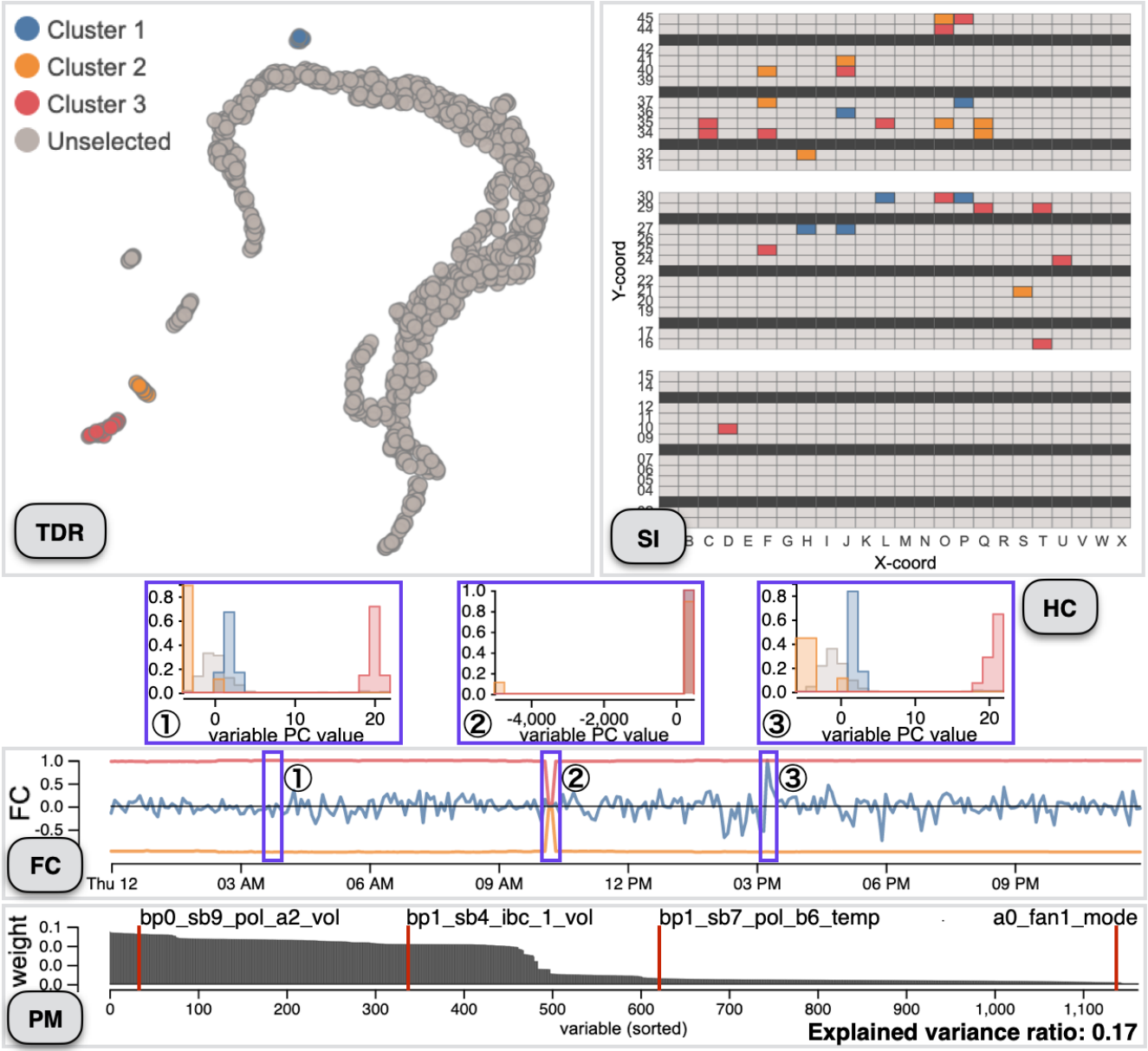}
    \caption{Case study 4-1. 
    (TDR) shows similarities of racks based on their temporal behaviors. 
    (SI) visualizes the racks' physical coordinates in the K computer.
    (FC, PM) are the FC and \PMView{}s after selecting three outliers from (TDR). 
    (HC) shows the \HCView{}s after selecting three different timestamps from (FC).}
    \label{fig:hw_log_1}
\end{figure}

\begin{figure}[tb]
    \centering
    \includegraphics[width=1.0\linewidth]{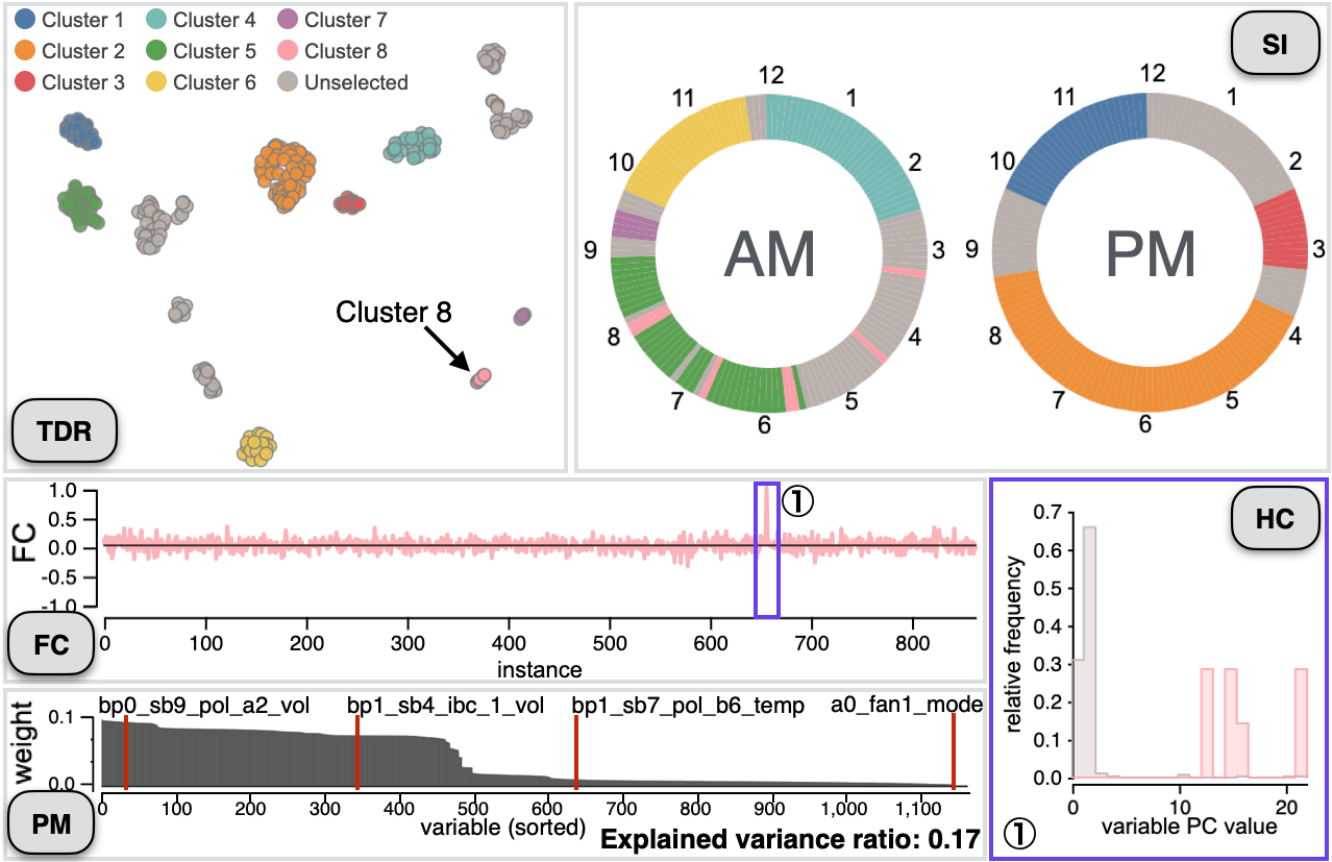}
    \caption{Case study 4-2. (TDR) shows similarities of timestamps based on behaviors of racks at the corresponding time. 
    (SI) informs the selected timestamps with a clock-based visualization.
    We show more information on Cluster 8 in the FC, PM, and \HCView{}s (FC, PM, HC).}
    \label{fig:hw_log_2}
\end{figure}

\vspace{2pt}
\noindent\textbf{Study 4-1: Identification and Characterization of Outlier Racks.}
As a first analysis, we identify racks that have unusual temporal behaviors.
To achieve this, we apply the two-step DR with PCA along a variable mode and then UMAP along a time mode (i.e., the dots in the TDR view represent instances). The visualized DR result is shown in \autoref{fig:hw_log_1}-TDR. 
We can see that while there is a large cluster that contains many racks (the gray points placed at the right side), some racks form small distinct clusters from the main cluster. 
We select three of these small clusters (Clusters 1--3) in \autoref{fig:hw_log_1}-TDR. 
Because these outlier clusters could relate to a specific physical location (e.g., a parallel application is often allocated to run in a specified location), we refer to the \SIView{} (\autoref{fig:hw_log_1}-SI), where the physical coordinates of racks are visualized; however, these clusters seem not to fit such a case.

To understand the clusters' characteristics, we analyze the results with the FC, HC, and \PMView{}s (\autoref{fig:hw_log_1}-FC, HC, PM). 
In \autoref{fig:hw_log_1}-FC, we first see that, across time, Clusters 1, 2, and 3 generally have moderate, strong negative, and strong positive FCs, respectively.
From \autoref{fig:hw_log_1}-FC, we select Timestamp  \textcircled{\small 1} as a sample timestamp following this general pattern and two timestamps (\textcircled{\small 2} and \textcircled{\small 3}) that have a unique shift of the FCs.
By looking at the \HCView{} of Timestamp \textcircled{\small 1}, we observe that Cluster 3 (red) has much higher feature values than the others, while Clusters 1 (blue) and 2 (orange) have slightly higher and lower feature values than the main cluster (gray), respectively.
However, at Timestamp \textcircled{\small 2}, all the racks have similar feature values.
At Timestamp \textcircled{\small 3}, when compared with Timestamp \textcircled{\small 1}, Cluster 1 has slightly less overlaps with the gray bins.
Next, we review the \PMView{} (\autoref{fig:hw_log_1}-PM), where 1,163 measures' weights are shown, and notice that only the first 500 measures have weights not close to zero. 
By showing these measures' names by hovering a mouse, we know that the first 500 measures are related to the voltages (``vol'') but not the others, including the temperatures (``temp'') and fan information (``fan\_mode'').

Therefore, we can conclude that, across time except for around 10~AM, the racks in Cluster 3 had extremely high voltages  while Cluster 1 and 2 had slightly higher and lower voltages than most of the racks.

\vspace{2pt}
\noindent\textbf{Study 4-2: Identification and Characterization of Outlier Timestamps.}
Next, we identify timestamps, at which racks had different behaviors from a usual state, by reviewing the two-step DR result generated by applying PCA along a variable mode and then UMAP along an instance mode (i.e., the dots in the TDR view represent timestamps). 
The visualized results are shown in \autoref{fig:hw_log_2}. 

From \autoref{fig:hw_log_2}-TDR, we select several distinct timestamp clusters (Clusters 1--8). 
In the \SIView{} (\autoref{fig:hw_log_2}-SI), where the corresponding timestamps are shown with a clock-based visualization, we can see most clusters relate to the specific time range (e.g., timestamps in Cluster 3 are seen from about 2 to 3~PM).
We can consider that each of these clusters corresponds to the duration when performing an allocated job. 
However, we can see that Cluster 8 is separated in several short time ranges in AM.
Since this pattern might relate to the abnormal behavior, we further review Cluster 8 by using the FC, PM, and \HCView{}s.
From \autoref{fig:hw_log_2}-FC, we can see that Cluster 8 has a strong positive FC for one instance (i.e., rack), as annotated with \textcircled{\small 1}. 
By looking at the \HCView{}, we can see that, for this instance, the timestamps belonging to Cluster 8 have a much higher feature value than the other timestamps (note: here the gray bins include all the timestamps except for those in Cluster 8).
Since the parametric mapping is the same as Study 4-1, we can say that these feature values mainly represent multiple voltage measures. 
Therefore, Cluster 8 is considered as outlier timestamps by the two-step DR because one specific rack had extremely high voltages at the corresponding timestamps.

\begin{figure*}[tb]
    \centering
    \includegraphics[width=0.955\linewidth]{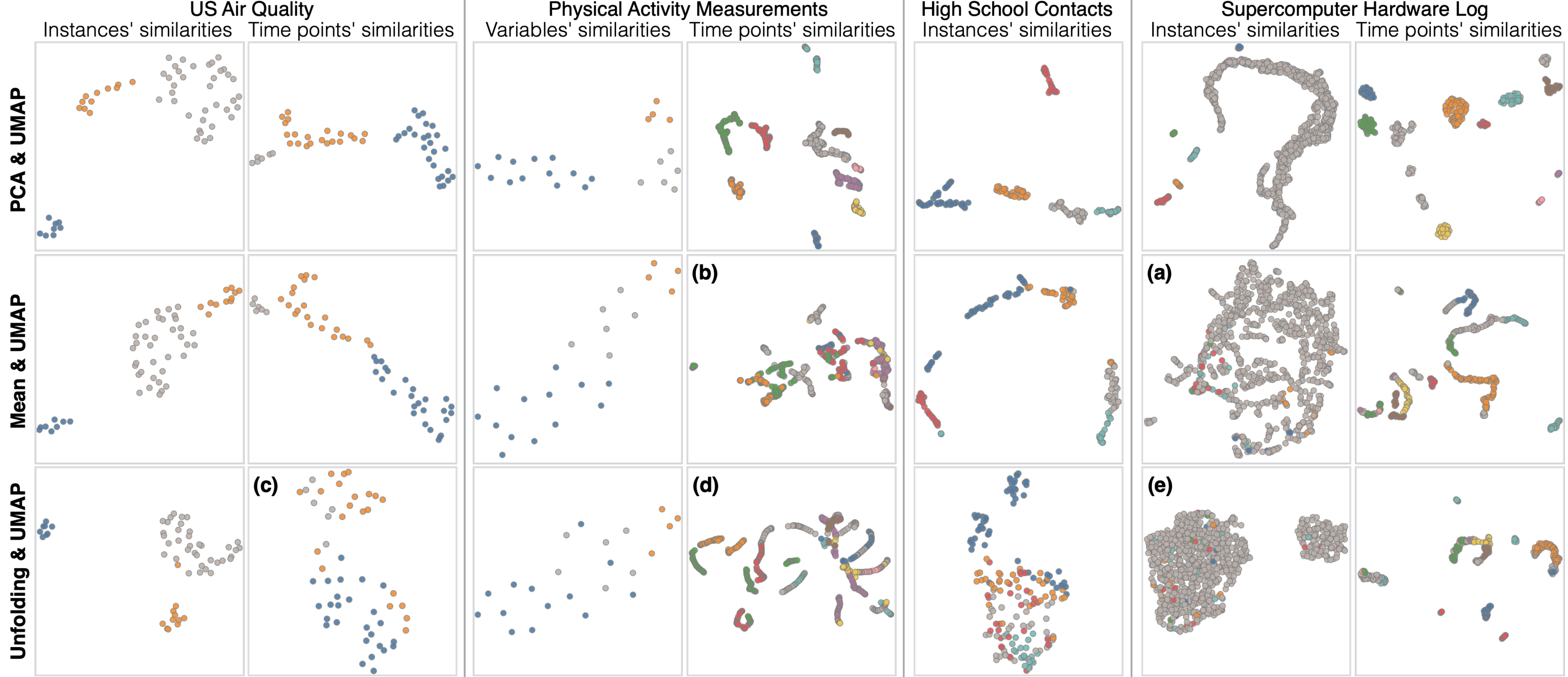}
    \caption{Visual comparison of the DR results. Colors represent clusters selected in each result of PCA \& UMAP.}
    \label{fig:qual_comp}
\end{figure*}

\section{Qualitative Comparison}
\label{sec:qual_comp}

The two-step DR in \name{} employs data compression with DR to produce a matrix from a third-order tensor.
Instead, as discussed in \autoref{sec:two_step_dr}, we can simply apply tensor unfolding along one mode to generate a matrix and then perform DR on such a matrix (e.g., applying DR on a matrix of $N$ rows and ($D \times T$) columns to visualize instance similarities). 
Another option is computing statistical measures, such as mean values, when generating a matrix from a third-order tensor. 
\name{} contains this approach if we consider the computation of statistical measures as one of DR methods that generate a representative value. 
Here, we compare three different methods above and discuss the advantages of the two-step DR.

\newcommand{\PU}{PCA \& UMAP}
\newcommand{\MU}{Mean \& UMAP}
\newcommand{\UU}{Unfolding \& UMAP}

More specifically, we compare two different implementations of the two-step DR, (1) using PCA for the first DR and UMAP for the second DR (we call this method \textit{\PU{}}) and (2) using the mean computation for the first DR and UMAP for the second DR (\textit{\MU{}}), and (3) the unfolding approach (i.e., without the first DR step) using UMAP as a DR method (\textit{\UU{}}). 
We apply these methods to the datasets analyzed in \autoref{sec:visual_interface} and \ref{sec:cs}. 

\autoref{fig:qual_comp} shows the DR results. 
Here, we manually select several distinct clusters from the results of \PU{} and then color-code the corresponding points in the other views based on the cluster information.
In general, some of the findings described in \autoref{sec:cs} from the results of \PU{} cannot be uncovered with either using \MU{} or \UU{}.

\vspace{2pt}
\noindent\textbf{\PU{} vs \MU{}.} 
\MU{} generates similar results with \PU{} when \PU{} generates a projection mapping consisting of
almost uniform weights (e.g., the results for the US air quality dataset); however, for the other cases, \MU{} does not show several meaningful clusters and outliers or does not clearly discern them from the other points. 
A concrete example can be seen in the instances' similarities of the supercomputer hardware log in \autoref{fig:qual_comp}-a, where the result of \MU{} mainly shows a single cluster and does not reveal the outlier clusters found with \PU{}.
Also, unlike \PU{}, the result in \autoref{fig:qual_comp}-b does not provide a clear separation of time points that are related to different physical activities.
When a target mode of the first DR has significant differences in variances for each index (i.e., a variable, an instance, or a time point), \PU{} can preserve more variety along the mode, and thus \PU{} would produce more useful results. 
However, again, the two-step DR does not restrict a DR method used for the first DR and allows the analyst to select a preferable compression/feature selection method, including the mean computation, PCA, LDA, etc. 

\vspace{2pt}
\noindent\textbf{\PU{} vs \UU{}.} 
\UU{} has quite different results from the ones with \PU{} and seems to fail to find several clusters and outliers. 
For example, in the time points' similarities of the US air quality dataset (\autoref{fig:qual_comp}-c), while \PU{} shows the clusters that represent the seasonal air quality change (as described in \autoref{sec:cs1}), \UU{} does not clearly display such clusters. Moreover, similar to \MU{}, the result in \autoref{fig:qual_comp}-d does not clearly discern different physical activities.
Also, \autoref{fig:qual_comp}-e does not uncover the outlier clusters seen in the result with \PU{}.
This limitation of \UU{} relates to the fact that \UU{} mixes two different modes together and, as a result, it cannot discover the patterns highly related to a specific mode. 
Another major drawback of \UU{} is that it makes the characterization of clusters more difficult because of the complexity of features in the \FCView{}, where each feature represents a mix of two modes (e.g., variables $\times$ instances), and the massiveness of the number of features (e.g., the supercomputer log dataset of $D=1,163$ and $N=864$ generates $D \times N = 1,004,832$ features).  

\section{Discussion}

We have evaluated \name{} with the case studies and qualitative comparison.
Through the qualitative comparison, we have discussed the strength of the two-step DR when compared with the other approaches. 
Here, we provide an additional discussion from different aspects.

\vspace{2pt}
\noindent\textbf{Limitations of Visual Scalability.}
In \name{}'s visual interface, we overlay multiple charts in the \FCView{} and the \HCView{} to make a comparison of different clusters' FCs and feature value distributions easier. 
However, when many clusters are selected (e.g., ten clusters), these visualizations could cause too many overlaps and clutters. 
To deal with such a situation, we can provide a visual comparison using small multiples and allow the analyst to select either overlays or small multiples based on their preference. 

Also, when each mode has many dimensions, it becomes difficult to grasp what kind of dimensions has high FCs and weights from the FC and \PMView{}s, respectively. 
This is especially problematic when the $x$-axis of the FC or \PMView{} represents a variable mode because it often consists of variables that have different types of measures (e.g., network routers' temperatures, voltages, sent, and received packets). 
For this issue, before visualizing the information of a variable mode, we can consider applying aggregation based on their similarities or available external information (e.g., a class of measures, such as physical loads, including temperatures and voltages, and network loads on routers, including sent and receive packets).  

\vspace{2pt}
\noindent\textbf{Limitations of the Two-Step DR.}
The two-step DR is mainly limited by the first DR step. 
Because this step compresses a target mode into 1D, when the mode has many dimensions (e.g., 1,000 variables in a variable mode), a large amount of information could be lost. 
However, at the same time, the analyst can check how much of the information is preserved by referring to a quality measure provided by each DR method, such as explained variance ratio in PCA and LDA. 
When the quality is extremely low (e.g., explained variance ratio is smaller than 0.01), the analyst can consider selecting a subset of dimensions for their analysis. 
In addition, to inform the second DR's quality, we plan to incorporate several model-agnostic quality measures~\cite{van2009dimensionality,lee2009quality} and  visualizations~\cite{wang2015revealing,fujiwara2017visual} in the future.

\vspace{2pt}
\noindent\textbf{Generality of the Two-Step DR.}
The back-end algorithms described in \autoref{sec:architecture} are used to obtain and understand a low-dimensional representation of multivariate time-series data. 
However, these algorithms can be applied to other types of data that can be formed into a third-order tensor.
For example, even when analyzing single-time-point multivariate data, the analyst may want to separate variables into two different modes, such as patients' demographics (e.g., ages) and their medical tests (e.g., blood pressures) for an analysis of medical datasets.
In such usage, \name{}'s algorithms can help the analyst avoid mixing the influences from two different types of variables on the DR result. 

\vspace{2pt}
\noindent\textbf{Additional Enhancement for Time-Series Analysis.}
Through this paper, we have demonstrated the effectiveness of \name{} using PCA, UMAP, and ccPCA as the first DR, second DR, and CL methods, respectively.
We also can use representation learning methods that focus on time-series analysis. 
For example, instead of PCA, when applying the first DR along a time mode, we are able to use functional PCA~\cite{wang2016functional}, which aims to extract representative temporal patterns, or use multivariate singular spectrum analysis (MSSA)~\cite{hassani2018multivariate}, which is suitable to find outlier time points. 
Also, we can design a CL method that is similar to ccPCA by extending a contrastive version of MSSA~\cite{dirie2019contrastive}.
Once it becomes available, we can replace ccPCA with such a method. 

Visualizations also can be enhanced for time-series analysis. 
For example, to convey the temporal order of time points in the \TDRView{}'s scatterplots, we can couple the \TDRView{} with the existing visualization methods described in \autoref{sec:related_work}, such as methods developed by Bach et al.~\cite{bach2016time} and 
van den Elzen et al.~\cite{van2016reducing}.

\section{Conclusion}

We have introduced a visual analytics framework, \name{}, which enables us to derive and interpret low-dimensional representations of multivariate time-series data by employing a two-step DR and contrastive learning together with interactive visualization.
As demonstrated with our case studies, \name{} can identify and characterize clusters and outliers from complex datasets.   
Therefore, \name{} provides a new effective approach to demanding tasks of analyzing multivariate time-series data.

\acknowledgments{This research is sponsored in part by the U.S. National Science Foundation through grant IIS-1741536 and U.S. Department of Energy through grant DE-SC0014917.}


\bibliographystyle{abbrv-doi}
\bibliography{00_main}
\end{document}